\documentclass[11pt]{article}
\usepackage{moriond,epsfig}

\def\Journal#1#2#3#4{{#1} {\bf #2}, #3 (#4)}

\def\PLB{{\em Phys. Lett.}  B}

\def\PRD{{\em Phys. Rev.} D}

\newcommand{\met}{\mbox{${\hbox{$E$\kern-0.6em\lower-.1ex\hbox{/}}}_T \!$ }}

\def\be{\begin{equation}}
\def\ee{\end{equation}}
\def\bea{\begin{eqnarray}}
\def\eea{\end{eqnarray}}

\begin{document}
\vspace*{4cm}
\title{$W$ BOSON MASS MEASUREMENT AT THE TEVATRON}

\author{ CHRISTOPHER P. HAYS \\ (for the CDF and D\O\ Collaborations)}

\address{Department of Physics, Duke University,
Durham, North Carolina 27708}

\maketitle\abstracts{
  The $W$ boson mass ($m_W$) is a key parameter of the standard model
(SM),
constraining the mass of the unobserved Higgs boson.  Using 
Tevatron $p\bar{p}$ collision data from 1992-1995, the 
CDF and D\O\ collaborations measured $m_W$ to $\delta m_W = 59$ MeV.  The 
ongoing Tevatron Run 2 has 
produced a factor of 5 more collisions, promising a significant 
reduction in $\delta m_W$.  CDF has analyzed
the first $\approx$200 pb$^{-1}$ of Run 2 data and determined its 
$\delta m_W$ to be 76 MeV.
}

\section{Introduction}
\label{sec:intro}

The SM describes all non-gravitational interactions
in terms of an $SU(3)_c \times SU(2)_L \times U(1)_Y$ gauge
symmetry.  Non-zero particle masses arise from the breaking of the 
$SU(2)_L \times U(1)_Y$ electroweak symmetry via the Higgs 
mechanism.  The Higgs boson is the last unobserved SM
particle, and the measured electroweak parameters severely
constrain its mass ($m_H$).  The constraint can be obtained from the
radiative correction $\Delta r$ to the $W$ boson mass ($m_W$) \cite{pdg}:
\begin{equation}
m_W^2 = \frac{\pi\alpha_{EM}}{\sqrt{2}G_F(1-m_W^2/m_Z^2)(1-\Delta r)}.
\end{equation}

\noindent
The correction $\Delta r \approx 0.67\%$ results predominantly from 
Higgs and $t\bar{b}$ loops in the $W$ boson propagator.  Because of
the precise measurements of the parameters $\alpha_{EM}$ 
($\delta {\alpha_{EM}}/\alpha_{EM}=0.014\%$ at $Q^2=m_Z^2$), $G_F$ 
($\delta {G_F}/G_F = 0.0009\%$), and $m_Z^2$ ($\delta {m_Z^2}/m_Z^2 
= 0.004\%$) \cite{pdg}, the uncertainties on $m_t$ and $m_W$ 
dominate the uncertainty on the inferred $m_H$.  To obtain equal 
$\chi^2$ contributions in a fit to $m_H$, the relation 
$\delta {m_W} = 0.007\delta {m_t}$
must hold \cite{bond}.  For the Run 1 $\delta {m_t}$ 
of 4.3 GeV, the required $\delta {m_W}$ is 30 MeV, close to 
$\delta {m_W}(world) = 34$ MeV \cite{pdg}.  The impending Run 2 
top mass measurements will significantly reduce $\delta {m_t}$, 
making $\delta {m_W}$ reduction of primary importance.
\par
The study of ongoing Run 2 $p\bar{p}$ collisions at the Tevatron will 
achieve this goal.  With 2 fb$^{-1}$ of data, the CDF and D\O\ 
collaborations expect to complete measurements with $\delta {m_W}$ between 
40 MeV \cite{cdf} and 50 MeV \cite{d0}.  Combining with the measurement from 
LEP ($\delta {m_W} = 42$ MeV) and the Run 2 $\delta {m_t} \approx 2$ GeV
will result in $\delta m_H/m_H \approx 30$\% \cite{lang}.
\par
The CDF and D\O\ collaborations are currently analyzing Run 2 data,
with D\O\ finalizing its event selection and precision calorimeter 
calibration, and CDF performing necessary cross-checks to its 
full analysis with $\approx$200 pb$^{-1}$ of data.  The CDF collaboration 
has determined the $W$ boson mass uncertainty associated with these data to be 
76 MeV.

\section{Measuring the $W$ Boson Mass at the Tevatron}

The $m_W$ measurement in $p\bar{p}$ data uses $s$-channel resonant
$W$ bosons with leptonic decays.  The transverse momentum of the decay 
$e$ or $\mu$ ($p_T^l$) can be measured with high precision and thus 
provides the bulk of the mass information.  Additional 
information comes from the decay $\nu$ transverse momentum ($p_T^{\nu}$), 
which is inferred from the measured energy imbalance in the event.  Since the 
lepton energy is well measured, the dominant uncertainty on $p_T^{\nu}$
comes from measuring the hadrons recoiling against the produced 
$W$ boson.  Because the $Z$ boson has a similar mass and production mechanism 
to the $W$ boson, events with $Z$ bosons can be used to calibrate and model 
the detector response to hadronic activity.
\par
The best statistical power for measuring $m_W$ is obtained by combining 
$p_T^l$ and $p_T^{\nu}$ into the transverse mass, defined as:

\begin{equation}
m_T = \sqrt{2 p_T^l p_T^{\nu}(1-\cos(\Delta\phi))}.
\label{eq:mt}
\end{equation}

\noindent
The transverse mass ignores the unmeasured $\nu$ momentum along
the beam direction ($\hat{z}$).  This distribution has a peak at 
$m_W$ (if we neglect detector resolution and final-state photon 
radiation) and a long tail below $m_W$, corresponding to events with 
$p_z^{\nu} \neq 0$.

\section{Run 2 CDF $W$ Boson Mass Measurement}

The relevant components of the CDF detector for the $m_W$ measurement 
are a large open-cell drift chamber immersed in a 1.4 T magnetic field, 
surrounded by a lead-scintillator sampling calorimeter.  Because of the 
similar resolutions and acceptances for 40 GeV $e$ and $\mu$, the combination 
of the two channels nearly doubles the 
effective statistics for the $m_W$ measurement. 
\par
The CDF strategy for the measurement proceeds as follows:  Model $W$ boson 
production and decay; 
calibrate track momentum using high-statistics resonances; calibrate 
calorimeter energy using $e$ tracks from $W$ boson decays; model hadronic 
response and resolution; estimate backgrounds; and fit the transverse
mass distribution to obtain $m_W$.

\subsection{Event Generation}

There are two important components of $W$ boson production for measuring 
$m_W$: the fractional momenta of $u$ and $d$ quarks inside 
the proton, and the $W~p_T$.  The $u$ and $d$  momenta 
determine $p_z^W$, which 
affects the $m_T$ distribution. The $u$ and $d$ fractional 
momenta are constrained from global fits to high-energy data and embodied 
in parton distribution functions (PDFs) independently parametrized by the 
CTEQ \cite{cteq} and MRST \cite{mrst} collaborations.  Using a CTEQ 
prescription for obtaining PDF uncertainties, the CDF 
collaboration has estimated $\delta m_W(PDF) = 15$ MeV. 
\par
\begin{figure}
\begin{minipage}{2.2in}
\epsfig{figure=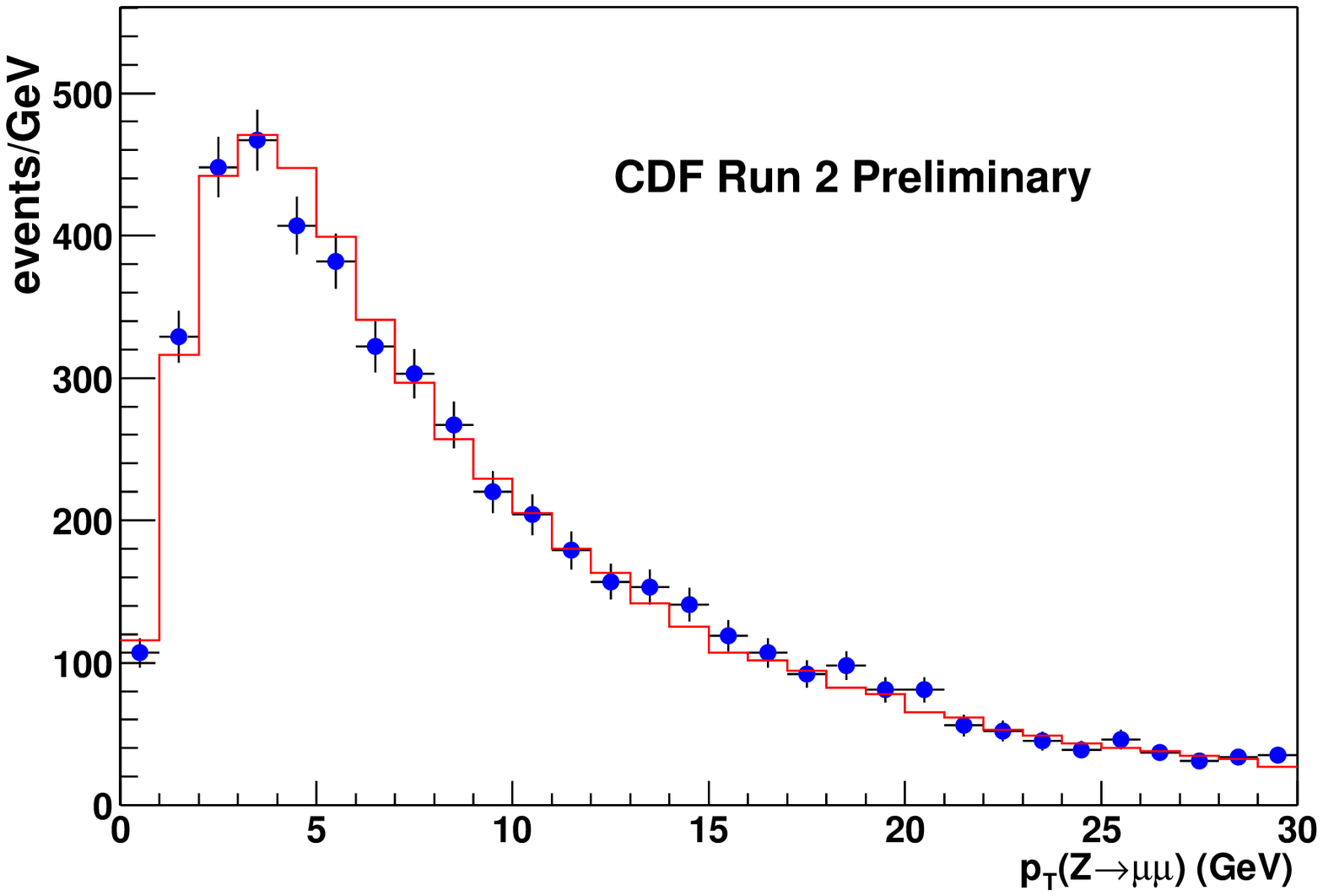,height=2.2in}
\end{minipage}
\hskip 1.in
\begin{minipage}{2.24in}
\epsfig{figure=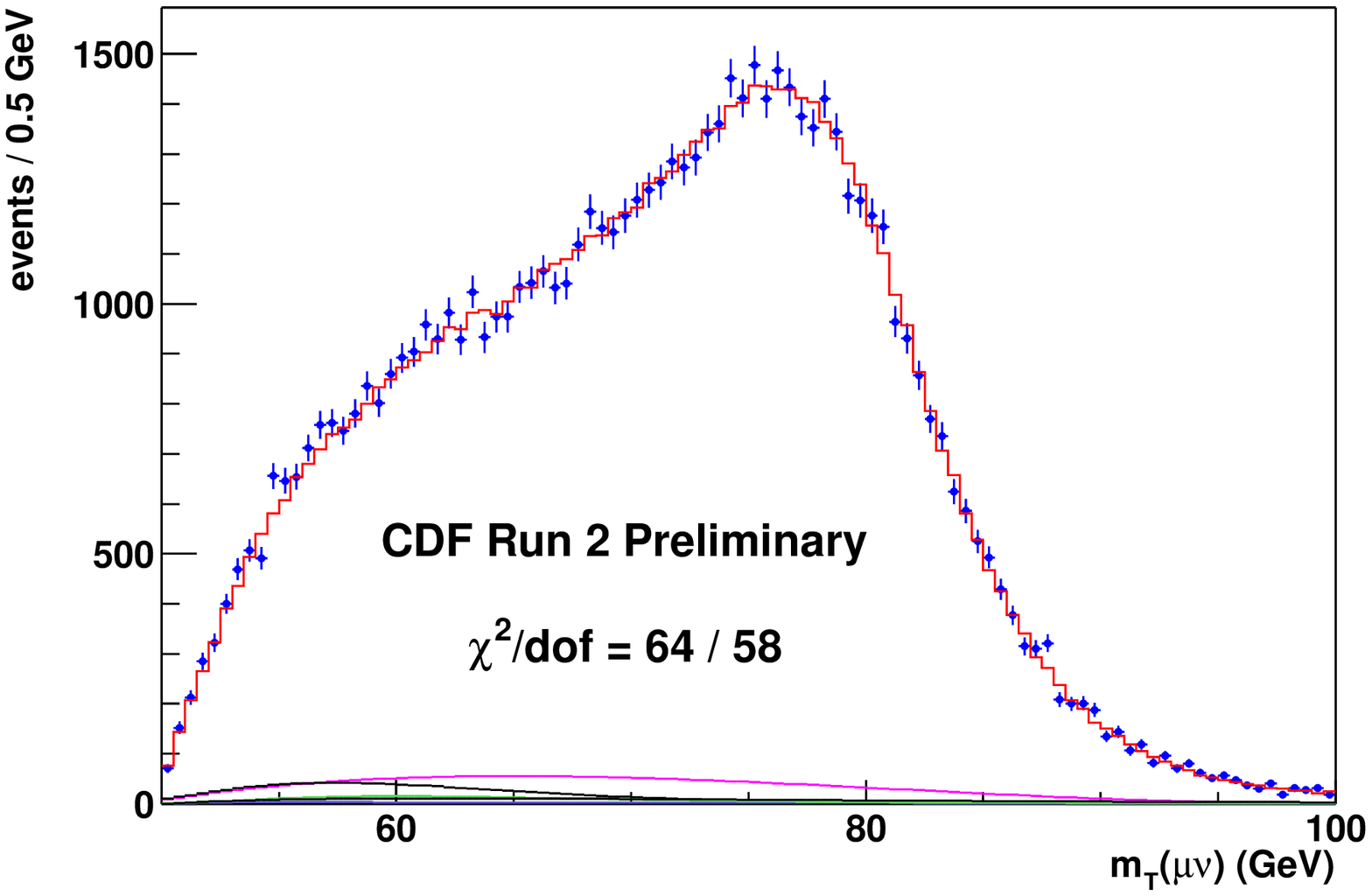,height=2.24in}
\end{minipage}
\caption{Left:  The $Z$ boson $p_T$ spectrum in CDF Run 2 $Z\rightarrow \mu\mu$
data (points) compared to the spectrum generated with {\sc resbos} 
(solid).  Right:  The $m_T$ distribution for $W$ boson decays to $\mu\nu$.
\label{fig:ptz}}
\end{figure}

The $W$ boson $p_T$ distribution is predicted by an event generator 
({\sc resbos} \cite{resbos}) that combines a QCD next-to-leading-log 
calculation with three non-perturbative parameters fit from high energy 
data.  The dominant constraint on these parameters comes from the $Z$ boson 
$p_T$ measurement in Run 1.  The generator and detector simulation predict 
the observed Run 2 $Z$ boson $p_T$ spectrum well (Fig. \ref{fig:ptz}).  The 
uncertainty on the {\sc resbos} parameters results in $\delta m_W(p_T^W) = 13$ 
MeV.

In the $W$ decay, the most important effect for the $W$ mass measurement 
is the radiation of a $\gamma$ from a final-state $l^{\pm}$.  This 
radiation results in a reduced $l^{\pm}$ momentum, potentially affecting the 
inferred mass of the $W$ boson.  CDF bases its simulation of final-state 
radiation on a QED next-to-leading order event generator 
({\sc wgrad}).  Effects from initial-state radiation, interference, and 
higher-order terms are not simulated, resulting in a 20 (15) MeV uncertainty 
for the $m_W$ measurement in the $\mu$ ($e$) channel.

\subsection{Track Momentum Calibration}

A charged particle's momentum is measured through its observed curvature 
in the tracker.  Since the momentum is inversely proportional to curvature, 
the momentum scale is measured as a function of the mean inverse momentum of 
$J/\psi$ muons and fit to a line.  The line has zero slope, verifying
the applicability of the extracted scale to $W$ boson decays.
\par
To improve momentum resolution, muon tracks from
$W$ and $Z$ decays use the beam position as a point in the track fit.
This constraint cannot be applied to $J/\psi$ decays since they can 
be separated from the beam line.  Instead, $\Upsilon$ decays are 
used to verify that the beam constraint produces no bias on the momentum 
calibration.  A systematic uncertainty of 15 MeV accounts 
for the observed difference in scale.  Including 
the uncertainty due to tracker alignment, CDF estimates
an uncertainty of $\delta m_W (p_T~scale) = 25$ MeV.  

\subsection{Calorimeter Energy Calibration}

Given the momentum calibration, electron tracks from $W$ decays are
used to calibrate the electromagnetic calorimeter.  The calorimeter
energy is scaled such that the ratio of energy to track momentum ($E/p$) 
is equal to 1.  To correct for an energy-dependent scale, the $E/p$ 
distribution is fit as a function of electron $E_T$ and a 
correction applied.  
\par
The significant amount of material in the silicon detector inside the 
tracker affects the position of the $E/p$ peak.  An uncertainty on the 
amount of material translates into an uncertainty on the measured $E$ scale.  
The fraction of events in the region $1.19 < E/p < 1.85$ is a measure
of the material.  The extent to which this region is not well modelled
results in a 55 MeV uncertainty on the $W$ mass.  This uncertainty dominates
the total $\delta m_W(E~scale)$ of 70 MeV.

\subsection{Hadronic Recoil Measurement and Simulation}

The hadronic recoil energy is measured by vectorially summing all the 
energy in the calorimeter, excluding that contributed by the $l$.
The detector response to the hadronic energy is defined as 
$R = u_{meas}/u_{true}$, where $u_{true}$ is the recoil energy of the 
$W$ boson.  The response is measured using $Z\rightarrow ll$, since 
the $l$ is measured more precisely than the hadronic energy.
\par
The hadronic energy resolution is modelled as having a component from
the underlying event (independent of recoil) and a component from the 
recoiling hadrons.  The model parameters are tuned using the resolution 
of $Z\rightarrow ll$ along the axis bisecting the leptons.  This 
axis is the least susceptible to fluctuations in $l$ energy.  
The recoil response and resolution uncertainty on the $W$ mass is
50 MeV, of which 37 MeV is due to the model of the underlying energy
resolution.  

\subsection{Backgrounds}

The backgrounds common to the $W\rightarrow e\nu$ and $W\rightarrow \mu\nu$ 
samples are: $Z\rightarrow ll$, where one $l$ is not reconstructed; 
$W\rightarrow \tau\nu\rightarrow l 3\nu$; and dijet production, with one 
hadronic jet misreconstructed as an $l$.  In addition, the $\mu$ sample
includes background from cosmic rays and decays in flight.  The $W$ and 
$Z$ backgrounds are estimated using Monte Carlo.  The dijet background 
estimation uses events with significant energy 
surrounding the $l$ to enhance hadronic background and obtain a 
background \met distribution.  The data \met distribution is then
fit using the $W$ and jet distributions as input.  The cosmic ray 
background is determined using track hit timing information and the
decay-in-flight background estimated by fitting the $\Delta \phi(l,\met)$ 
distribution to a combination of $W$ and decay-in-flight distributions.
These estimates result in $\delta m_W(background) = 20$ MeV.

\subsection{Mass Fit and Systematics}

Given the energy calibrations, recoil model, and background estimation, 
the $m_T$ distribution is fit for the $e$ and $\mu$
channels.  The predicted line shape agrees with that of the data (Fig. 1).  The
central value is blinded while CDF cross-checks the analysis with independent
data sets and simulation.  Combining the two channels (Table 1) 
results in $\delta m_W = 76$ MeV.

\begin{table}[t]
\caption{The uncertainties on the $W$ boson mass measurement in MeV/$c^2$ 
using 0.2 fb$^{-1}$ of Run 2 CDF data.  The CDF Run 1B
uncertainties are shown for comparison.
}\label{uncert}
\vspace{0.4cm}
\begin{center}
\begin{tabular}{|c|c|c|}
\hline
Sytematic Uncertainty & Electrons (Run 1B \cite{1b}) & Muons (Run 1B \cite{1b}) \\
\hline
Production and Decay Model & 30 (30) & 30 (30) \\
Lepton $E$ Scale and Resolution & 70 (80) & 30 (87) \\
Recoil Scale and Resolution & 50 (37) & 50 (35) \\
Backgrounds & 20 (5) & 20 (25) \\
Statistics & 45 (65) & 50 (100) \\
\hline
Total & 105 (110) & 85 (140) \\
\hline
\end{tabular}
\end{center}
\end{table}

\end{document}